\PassOptionsToPackage{unicode}{hyperref}
\PassOptionsToPackage{hyphens}{url}
\PassOptionsToPackage{dvipsnames,svgnames,x11names}{xcolor}
\documentclass[
  12pt]{article}

\usepackage{amsmath,amssymb}
\usepackage{iftex}
\ifPDFTeX
  \usepackage[T1]{fontenc}
  \usepackage[utf8]{inputenc}
  \usepackage{textcomp} % provide euro and other symbols
\else % if luatex or xetex
  \usepackage{unicode-math}
  \defaultfontfeatures{Scale=MatchLowercase}
  \defaultfontfeatures[\rmfamily]{Ligatures=TeX,Scale=1}
\fi
\usepackage{lmodern}
\ifPDFTeX\else  
\fi
\IfFileExists{upquote.sty}{\usepackage{upquote}}{}
\IfFileExists{microtype.sty}{% use microtype if available
  \usepackage[]{microtype}
  \UseMicrotypeSet[protrusion]{basicmath} % disable protrusion for tt fonts
}{}
\makeatletter
\@ifundefined{KOMAClassName}{% if non-KOMA class
  \IfFileExists{parskip.sty}{%
    \usepackage{parskip}
  }{% else
    \setlength{\parindent}{0pt}
    \setlength{\parskip}{6pt plus 2pt minus 1pt}}
}{% if KOMA class
  \KOMAoptions{parskip=half}}
\makeatother
\usepackage{xcolor}
\makeatletter
\ifx\paragraph\undefined\else
  \let\oldparagraph\paragraph
  \renewcommand{\paragraph}{
    \@ifstar
      \xxxParagraphStar
      \xxxParagraphNoStar
  }
  \newcommand{\xxxParagraphStar}[1]{\oldparagraph*{#1}\mbox{}}
  \newcommand{\xxxParagraphNoStar}[1]{\oldparagraph{#1}\mbox{}}
\fi
\ifx\subparagraph\undefined\else
  \let\oldsubparagraph\subparagraph
  \renewcommand{\subparagraph}{
    \@ifstar
      \xxxSubParagraphStar
      \xxxSubParagraphNoStar
  }
  \newcommand{\xxxSubParagraphStar}[1]{\oldsubparagraph*{#1}\mbox{}}
  \newcommand{\xxxSubParagraphNoStar}[1]{\oldsubparagraph{#1}\mbox{}}
\fi
\makeatother

\usepackage{longtable,booktabs,array}
\usepackage{calc} % for calculating minipage widths
\usepackage{etoolbox}
\makeatletter
\patchcmd\longtable{\par}{\if@noskipsec\mbox{}\fi\par}{}{}
\makeatother
\IfFileExists{footnotehyper.sty}{\usepackage{footnotehyper}}{\usepackage{footnote}}
\makesavenoteenv{longtable}
\usepackage{graphicx}
\makeatletter
\def\maxwidth{\ifdim\Gin@nat@width>\linewidth\linewidth\else\Gin@nat@width\fi}
\def\maxheight{\ifdim\Gin@nat@height>\textheight\textheight\else\Gin@nat@height\fi}
\makeatother
\setkeys{Gin}{width=\maxwidth,height=\maxheight,keepaspectratio}
\makeatletter
\def\fps@figure{htbp}
\makeatother

\makeatletter
\@ifpackageloaded{caption}{}{\usepackage{caption}}
\AtBeginDocument{%
\ifdefined\contentsname
  \renewcommand*\contentsname{Table of contents}
\else
  \newcommand\contentsname{Table of contents}
\fi
\ifdefined\listfigurename
  \renewcommand*\listfigurename{List of Figures}
\else
  \newcommand\listfigurename{List of Figures}
\fi
\ifdefined\listtablename
  \renewcommand*\listtablename{List of Tables}
\else
  \newcommand\listtablename{List of Tables}
\fi
\ifdefined\figurename
  \renewcommand*\figurename{Figure}
\else
  \newcommand\figurename{Figure}
\fi
\ifdefined\tablename
  \renewcommand*\tablename{Table}
\else
  \newcommand\tablename{Table}
\fi
}
\@ifpackageloaded{float}{}{\usepackage{float}}
\floatstyle{ruled}
\@ifundefined{c@chapter}{\newfloat{codelisting}{h}{lop}}{\newfloat{codelisting}{h}{lop}[chapter]}
\floatname{codelisting}{Listing}

\makeatother
\makeatletter
\@ifpackageloaded{subcaption}{}{\usepackage{subcaption}}
\makeatother

\ifLuaTeX
  \usepackage{selnolig}  % disable illegal ligatures
\fi
\usepackage[]{natbib}
\usepackage{bookmark}

\IfFileExists{xurl.sty}{\usepackage{xurl}}{} % add URL line breaks if available
\hypersetup{
  pdftitle={A Semiparametric Functional Generalized Linear Model for Multi-Population Data},
  pdfauthor={},
  pdfkeywords={density ratio model, empirical likelihood, functional data analysis, penalized B-splines, semiparametric regression},
  colorlinks=true,
  linkcolor={blue},
  filecolor={Maroon},
  citecolor={Blue},
  urlcolor={Blue},
  pdfcreator={LaTeX}}

\newcommand{\anon}{1}

% ...
% ...
\usepackage{epsfig,fancyheadings,rotating}
\usepackage{mathrsfs}
\usepackage{bm,bbm}
\usepackage{sectsty, secdot}
\usepackage{amsfonts}
\usepackage{multirow}
\usepackage{amsthm}
\usepackage[ruled,linesnumbered,lined,commentsnumbered]{algorithm2e}
\usepackage{mathtools}
\usepackage{enumitem}
\usepackage{threeparttable}

% ...
% ...
% ...
\newcommand{\be}{\begin{equation}}
\newcommand{\ee}{\end{equation}}
\newcommand{\bit}{\begin{itemize}}
\newcommand{\eit}{\end{itemize}}

\newcommand{\bas}{\begin{eqnarray*}}
\newcommand{\eas}{\end{eqnarray*}}
\newcommand{\ba}{\begin{eqnarray}}
\newcommand{\ea}{\end{eqnarray}}

\newcommand{\bgamma}{\mbox{\boldmath $\gamma$}}

\newcommand{\btheta}{\mbox{\boldmath $\theta$}}

\newtheorem{theorem}{Theorem}

\theoremstyle{definition}

\begin{document}

\def\spacingset#1{\renewcommand{\baselinestretch}%
{#1}\small\normalsize} \spacingset{1}

%%%%%%%%%%%%%%%%%%%%%%%%%%%%%%%%%%%%%%%%%%%%%%%%%%%%%%%%%%%%%%%%%%%%%%%%%%%%%%

\if1\anon
{
  \title{\bf A Semiparametric Functional Generalized Linear Model for Multi-Population Data}
  \author{Boyi Hu\hspace{.2cm}\\
    Department of Neurology, Columbia University\\
    and \\
    Xixi Hu \\
   Sauder School of Business, University of British Columbia\\
   and \\
   Jiguo Cao\\
   Department of Statistics and Actuarial Science, Simon Fraser University
   }
  \maketitle
} \fi

\if0\anon
{
  \bigskip
  \bigskip
  \bigskip
  \begin{center}
    {\LARGE\bf A Semiparametric Functional Generalized Linear Model for Multi-Population Data}
\end{center}
  \medskip
} \fi

\bigskip
\begin{abstract}
Functional generalized linear models provide a flexible framework for relating a scalar response to functional and scalar predictors, but their conventional formulation requires specification of the response distribution and is typically developed for a single population. We propose a semiparametric functional generalized linear model for multi-population data that leaves the baseline response distributions unspecified and links them across populations through a density ratio model. The proposed framework accommodates both functional and scalar predictors while borrowing information across related populations. We develop a maximum empirical likelihood estimation procedure and use penalized B-spline approximations to estimate the functional coefficients. A cross-validation procedure based on a kernel-smoothed empirical likelihood density estimator is introduced to select the smoothing parameters. Simulation studies show that borrowing information through the density ratio structure can improve estimation efficiency relative to fitting each population separately, and that the proposed semiparametric method remains competitive when a parametric Gaussian model is correctly specified while providing substantial robustness when it is misspecified. We illustrate the method using county-level soybean yield data from Kansas, with daily temperature curves and irrigation levels as predictors.
\end{abstract}

\noindent%

\vfill

\newpage
\spacingset{1.8} % JCGS template body spacing.

\section{Introduction}
Soybean holds significant global importance because of its vital role in meeting food, feed, and fuel demands. A substantial proportion, approximately 77\%, of global soy production is used for livestock feed to support meat and dairy production. The remaining soy is used for a variety of purposes, including fuel, industrial applications, and vegetable oils, while a smaller proportion, around 7\%, is directly consumed by humans in forms such as tofu and soy milk.
The germination and growth of soybeans heavily depend on environmental resources, particularly temperature and water availability \citep{suo2022soybean}. Consequently, these environmental factors can have a substantial impact on soybean yield at the end of the growing season. To study the relationship between soybean yield and its environmental determinants, we collected county-level soybean yield data together with two related predictors, daily temperature and irrigation level, in Kansas over the period from 1993 to 2004. By analyzing these data, we aim to gain insight into how temperature and water resources influence final soybean yield. Such information may be useful for researchers and practitioners seeking to better understand soybean production under varying environmental conditions.

To investigate these relationships, we propose a novel semiparametric functional generalized linear model (FGLM) for the relationship between county-level soybean yield, considered as a continuous response variable, a scalar predictor describing irrigation level, and a functional predictor consisting of daily temperature measurements. The proposed model provides a flexible and interpretable representation of the association between the response and predictors while allowing the response distributions to vary across multiple populations.

The functional generalized linear model (FGLM) \citep{ramsayfunctional, james2002generalized} extends the generalized linear model (GLM) \citep{mccullagh2019generalized} by directly modeling the relationship between a scalar response variable, typically assumed to follow a member of the exponential family, and a functional predictor. For a random variable $Y$ with density
\[
p(y;\eta,\phi)=\exp\left\{\frac{y\eta - b(\eta)}{a(\phi)} + c(y,\phi)\right\},
\]
the FGLM models the relationship between a functional predictor $X(t)$ and response $Y$ as
\be
h(\mu) = \beta_0 + \int_T X(t)\beta(t)dt,
\label{fglm}
\ee
where $\mu = \mathbbm E(Y|\eta,\phi) = b'(\eta)$ and $h(\cdot)$ is a link function. Depending on the distribution of $Y$, different link functions are used. For example, when $Y$ follows a Gaussian distribution, the identity link is commonly used, whereas for a binary response the logit link is typically adopted.

Model \eqref{fglm} is a valuable tool in environmental applications, particularly when functional observations such as daily temperature or humidity are available. However, several challenges arise in real applications. One important concern is the need to specify the functions $a(\phi)$, $b(\eta)$, and $c(y,\phi)$ so that $\exp\{c(y,\phi)\}$ corresponds to a valid baseline distribution. If the underlying response distribution is misspecified, estimation of the functional coefficient $\beta(t)$ in model \eqref{fglm} can be affected, potentially reducing the reliability and interpretability of the fitted model.
A second challenge is that agricultural data sets are often collected across multiple years or locations. Samples from different years or locations may be regarded as arising from distinct but related populations because of differences in environmental conditions, management practices, and other unmeasured factors. Directly fitting model \eqref{fglm} to each sample separately may therefore be inefficient, particularly when the sample size within each population is limited. This issue is especially relevant in our soybean application. Important environmental factors such as humidity and sunshine are not available in the data, so the response distributions may vary across years because of unmeasured annual conditions. In addition, most years between 1993 and 2004 contain only approximately 20--40 county-level observations, which makes separate estimation for each year challenging under semiparametric models.

To address these issues, we first consider a semiparametric functional generalized linear model with an unspecified baseline distribution. Let $\bm Z$ be a vector of scalar predictors and $X(t)$ be a functional predictor. For a single population, we propose
\be
f(y|\bm z, x(t))=\frac{\text{exp}\left\{\left(\bm z^\top\bgamma + \langle x, \beta\rangle\right) y\right\}g(y)}{\int \text{exp}\left\{\left(\bm z^\top\bgamma + \langle x, \beta\rangle\right) y\right\}g(y)dy},
\label{fprop}
\ee
where $g(y)$ is the density of an unknown baseline distribution $G(y)$ and $f(y|\bm z,x(t))$ denotes the conditional density of $Y$ given $\bm Z=\bm z$ and $X(t)=x(t)$. In model \eqref{fprop}, the vector $\bgamma$, the function $\beta(t)$, and the density $g(y)$ are all unknown, and $\langle x,\beta\rangle=\int_0^T\beta(t)x(t)dt$ denotes the inner product between $x(t)$ and $\beta(t)$. By leaving $g(y)$ unspecified, model \eqref{fprop} reduces reliance on a fully specified parametric response distribution and therefore provides greater robustness to distributional misspecification than the conventional formulation in \eqref{fglm}.

To further address the limited sample size within individual populations, we incorporate the density ratio model (DRM) \citep{anderson1979multivariate,chen2013quantile, zhang2022density}. Suppose we have $m+1$ populations with distribution functions $G_r(y)$ and density functions $g_r(y)$, $r=0,\ldots,m$. For instance, $G_r(y)$ represents the baseline distribution of county-level soybean yield in Kansas for year $r$. The DRM links the baseline densities through
\be
\frac{g_r(y)}{g_0(y)} = \exp\{\alpha_r + \bm\theta_r^\top \bm q(y)\},\label{drm}
\ee
where $\bm q(y)$ is a vector of prespecified basis functions and $\alpha_r$ is a normalizing constant. The flexibility of the DRM is an important advantage. For example, choosing $\bm q(y)=(y,y^2)^\top$ accommodates Normal distribution families, whereas $\bm q(y)=(y,\log y)^\top$ accommodates Gamma distribution families. In practice, a richer vector $\bm q(y)$ can be considered, with model-selection criteria such as AIC or BIC used to select a suitable specification. Detailed discussions of density ratio models for linking multiple populations can be found in \cite{anderson1979multivariate, qin1997goodness, chen2013quantile, zhang2022density}.

Combining \eqref{fprop} and \eqref{drm}, we focus on the following semiparametric functional generalized linear model:
\begin{align}
f_k(y|\bm z, x(t))&=\frac{\text{exp}\left\{\left(\bm z^\top\bgamma_k + \langle x, \beta_k\rangle\right) y\right\}g_k(y)}{\int \text{exp}\left\{\left(\bm z^\top\bgamma_k + \langle x, \beta_k\rangle\right) y\right\}g_k(y)dy},\quad k=0,\ldots, m,\label{fprop_drm}\\
\frac{g_r(y)}{g_0(y)} &= \exp\{\alpha_r + \bm\theta_r^\top \bm q(y)\},\quad r=1,\ldots, m.\nonumber
\end{align}
Here, $f_k(y|\bm z,x(t))$ is the conditional density of $Y$ and $g_k(y)$ is the density of an unknown baseline distribution $G_k(y)$ for population $k$. The baseline densities $(g_0,\ldots,g_m)$ are linked through the DRM structure. This framework has two main advantages over a conventional FGLM fitted separately to each population. First, the baseline distributions are left unspecified, reducing sensitivity to parametric distributional assumptions. Second, the DRM structure permits simultaneous estimation using pooled data, allowing related populations to borrow information while retaining population-specific regression parameters.

\subsection{Interpretation of the proposed multi-population model}
To better understand model \eqref{fprop_drm}, define the population-specific linear predictor
\[
\eta_k(\bm z,x)=\bm z^\top\bgamma_k+\langle x,\beta_k\rangle.
\]
Conditional on the predictors, the baseline density $g_k(y)$ is exponentially tilted by the factor $\exp\{\eta_k(\bm z,x)y\}$. Thus, the regression parameters do not directly specify a conditional mean as in an ordinary linear regression model. Instead, they determine how the entire conditional response distribution is tilted relative to the baseline distribution. For two predictor values with linear predictors $\eta_{k,1}$ and $\eta_{k,2}$, the corresponding conditional densities satisfy, up to their normalizing constants,
\[
\frac{f_k(y\mid \eta_{k,2})}{f_k(y\mid \eta_{k,1})}
\propto \exp\{(\eta_{k,2}-\eta_{k,1})y\}.
\]
Consequently, an increase in the linear predictor places relatively more weight on larger response values when $\eta_{k,2}>\eta_{k,1}$ and on smaller response values when $\eta_{k,2}<\eta_{k,1}$. This distributional interpretation is useful for both the scalar coefficient $\bgamma_k$ and the functional coefficient $\beta_k(t)$.
The proposed formulation also contains familiar parametric models as special cases. For example, if the baseline distribution $G_k$ is Gaussian with variance $\sigma_k^2$, exponential tilting preserves the Gaussian family and shifts its conditional mean by an amount proportional to $\sigma_k^2\eta_k(\bm z,x)$. Therefore, after a rescaling of the regression coefficients, model \eqref{fprop_drm} includes the usual Gaussian functional linear model. The key distinction is that the proposed method does not require the analyst to assume a Gaussian, Gamma, or another fully specified baseline distribution in advance.

The multi-population structure separates two sources of heterogeneity. The regression parameters $(\bgamma_k,\beta_k)$ are population-specific, so the effects of scalar and functional predictors are allowed to differ across populations. At the same time, the baseline distributions $G_k$ are also permitted to vary. The DRM in \eqref{drm} does not force these baseline distributions to be identical; it only assumes that their density ratios can be represented through a finite-dimensional exponential tilt. This provides a middle ground between fitting each population completely separately and imposing a common response distribution across all populations. Information is shared through the linked baseline distributions, while population-specific regression relationships are retained.
For a single population, model \eqref{fprop} can be viewed as a functional extension of the proportional likelihood ratio model (PLRM) proposed in prior literature \citep{luo2012proportional},
\be
f(y|\bm z)=\frac{\text{exp}\left\{\bm z^\top\bm\beta y\right\}g(y)}{\int \text{exp}\left\{\bm z^\top\bm\beta y\right\}g(y)dy}.
\label{prop_ratio}
\ee
Model \eqref{prop_ratio} provides a parsimonious way to assess nonlinear monotone relationships between a scalar response and finite-dimensional predictors. Its connections with GLMs \citep{mccullagh2019generalized}, density ratio models \citep{qin1997goodness}, biased sampling models \citep{gilbert1999maximum}, single-index models \citep{ichimura1993semiparametric}, and exponential tilt regression models \citep{rathouz2009generalized} are discussed by \cite{luo2012proportional}. The PLRM has also been extended to a variety of applications, including engineering, biomedical research, and survival analysis \citep{luo2015moment,huang2012proportional,huang2013semiparametric,ning2017likelihood}.

The main contributions of this paper are threefold. First, we extend the semiparametric proportional likelihood ratio framework to accommodate functional predictors and multiple related populations through a density ratio structure. Second, we develop a maximum empirical likelihood estimation procedure with penalized B-spline approximation for population-specific functional coefficients, together with a cross-validation strategy for selecting smoothing parameters. Third, we evaluate the proposed method through simulation studies under both correctly specified and misspecified parametric response distributions and illustrate its practical utility using multi-year soybean yield data from Kansas.

\section{Estimation}
\subsection{Empirical likelihood based on the proposed model}
Suppose the baseline population distribution functions, $G_k(y)$, of $m+1$ samples with sample sizes $n_k$ satisfy
\be
\frac{g_k(y)}{g_0(y)} = \exp\{\alpha_k + \bm\theta_k^\top\bm q(y)\} \quad k=1,\ldots, m,
\label{DRM}
\ee
where $g_k(y)$ denotes the density function of $G_k(y)$.
For notational simplicity, we set $\alpha_0=0$ and $\btheta_0 = \bm 0$.
Assume i.i.d. data $\left\{\bm z_{k,i}, x_{k,i}(t), y_{k,i}\right\}$, $k=0,\ldots,m; i=1,\ldots,n_k$ from distributions that satisfy the model,
\be
f_k(y|\bm z, x(t))=\frac{\text{exp}\left\{\left(\bm z^\top\bgamma_k + \langle x, \beta_k\rangle\right) y\right\}g_k(y)}{\int \text{exp}\left\{\left(\bm z^\top\bgamma_k + \langle x, \beta_k\rangle\right) y\right\}g_k(y)dy},\quad k=0,\ldots, m.\label{FPROP}
\ee
We use the maximum empirical likelihood (EL) method to fit the model \eqref{FPROP}.
Empirical likelihood methods under the DRM and PLRM have been studied by \cite{fokianos2001semiparametric,qin1997goodness,luo2012proportional}.
For convenience, we first introduce some notation.
Define
\[
A(\bm z, x(t);\bgamma, \beta(t),G(y)) = \int \text{exp}\left\{\left(\bm z^\top\bgamma + \langle x, \beta\rangle\right)y\right\}dG(y).
\]
The contribution to the likelihood function of one observation from population $k$, $(\bm z_{k,i}, x_{k,i}(t), y_{k,i})$ is given by 
\[
lik(\bgamma_k, \beta_k,G_k) = \text{exp}\left\{\left(\bm z_{k,i}^\top\bgamma_k + \langle x_{k,i}, \beta_k\rangle\right)y_{k,i}\right\} \Delta G_k(y_{k,i})/A(\bm z_{k,i}, x_{k,i}(t);\bgamma_k, \beta_k(t), G_k),
\]
where $\Delta G_k(y) = G_k(y+)-G_k(y-)$.
Then the log-likelihood for $(\bgamma_k, \beta_k,G_k)$ is
\begin{align*}
\sum_{i=1}^{n_k}\left\{\bm z_{k,i}^\top\bgamma_k + \langle x_{k,i}, \beta_k\rangle\right\} y_{k,i} + \sum_{i=1}^{n_k}\log\Delta G_k(y_{k,i})- \sum_{i=1}^{n_k} \log A(\bm z_{k,i}, x_{k,i}(t);\bgamma_k, \beta_k(t),G_k).
\end{align*}
Under the DRM assumption on the baseline distributions $G_k(y)$, the log-likelihood function of $(\bgamma_k, \beta_k,G_k)$ can also be viewed as a function of $(\bgamma_k, \beta_k,\alpha_k,\btheta_k, G_0)$.
Then the log-likelihood function of $(\bgamma_k, \beta_k,\alpha_k,\btheta_k, G_0)$ can be written as
\begin{align}
l_n(\bgamma_k, \beta_k,\alpha_k, \btheta_k, G_0)=& \sum_{i=1}^{n_k}\left\{\bm z_{k,i}^\top\bgamma_k + \langle x_{k,i}, \beta_k\rangle\right\} y_{k,i} + \sum_{i=1}^{n_k}\log\Delta G_0(y_{k,i})\nonumber\\
&+ \sum_{i=1}^{n_k}\left\{\alpha_k + \bm q(y_{k,i})^\top\btheta_k\right\}
- \sum_{i=1}^{n_k} \log A(\bm z_{k,i}, x_{k,i}(t);\bgamma_k, \beta_k,G_k).\label{initial_el}
\end{align}
From the empirical likelihood point of view, the distributions $G_k(y)$ can be treated as discrete distributions.
Let $p_{k,i}$ denote the mass that the distribution function $G_0(y)$ assigns at $y_{k,i}$.
Then the log-likelihood function \eqref{initial_el} can be further written as
\begingroup
\footnotesize % Fit the long likelihood display within the JCGS text block.
\begin{align*}
l_n(\bgamma_k, \beta_k,\alpha_k, \btheta_k, G_0)=& \sum_{i=1}^{n_k}\left\{\bm z_{k,i}^\top\bgamma_k + \langle x_{k,i}(t), \beta_k(t)\rangle\right\} y_{k,i} + \sum_{i=1}^{n_k}\log p_{k,i}\\
&+ \sum_{i=1}^{n_k}\left\{\alpha_k + \bm q(y_{k,i})^\top\btheta_k\right\}\\
& - \sum_{i=1}^{n_k} \log\left(\sum_{r=0}^m\sum_{j=1}^{n_r}p_{r,j}\text{exp}\left\{\left(\bm z_{k,i}^\top\bgamma_k +
\langle x_{k,i}(t), \beta_k(t)\rangle \right)y_{r,j} + \alpha_k + \bm q(y_{r,j})^\top\btheta_k\right\} \right),
\end{align*}
\endgroup
under the constraints
\begin{align*}
&\sum_{k=0}^m\sum_{i=1}^{n_k} p_{k,i} = 1,\\
&\sum_{k=0}^m\sum_{i=1}^{n_k} \exp\{\alpha_r + \bm q(y_{k,i})^\top\btheta_r\}p_{k,i} = 1, \quad r=1,\ldots,m,\\
& p_{k,i}>0, \quad k=0,\ldots,m, \quad i=1,\ldots, n_k.
\end{align*}
Let 
$\bgamma =(\bgamma_0^\top, \ldots, \bgamma_m^\top)^\top$,
$\bm \beta = (\beta_0(t), \ldots, \beta_m(t))^\top$,
$\bm\alpha = (\alpha_1,\ldots,\alpha_m)^\top$,
$\btheta = (\btheta_1^\top,\ldots,\btheta_m^\top)^\top$,
and $\bm p =\{p_{k,i}\}^\top_{k=0,\ldots,m;i=1,\ldots,n_k}$.
Then the log-likelihood function of $(\bgamma, \bm \beta, \bm\alpha,\btheta, \bm p)$ is given by
\begin{align}
l_n(\bgamma,\bm\beta,\bm\alpha,\btheta,\bm p)
={}&
\sum_{k=0}^m\sum_{i=1}^{n_k}
\left(\bm z_{k,i}^\top\bgamma_k
+\langle x_{k,i}(t),\beta_k(t)\rangle\right)y_{k,i}
\nonumber\\
&+\sum_{k=0}^m\sum_{i=1}^{n_k}\log p_{k,i}
+\sum_{k=0}^m\sum_{i=1}^{n_k}
\bigl\{\alpha_k+\bm q(y_{k,i})^\top\btheta_k\bigr\}
\nonumber\\
&-\sum_{k=0}^m\sum_{i=1}^{n_k}
\log\biggl(
\sum_{r=0}^m\sum_{j=1}^{n_r}p_{r,j}
\exp\Bigl\{
\left(\bm z_{k,i}^\top\bgamma_k
+\langle x_{k,i}(t),\beta_k(t)\rangle\right)y_{r,j}
\nonumber\\
&\hspace{4cm}
+\alpha_k+\bm q(y_{r,j})^\top\btheta_k
\Bigr\}\biggr),
\label{loglike0}\\
&\text{subject to}\nonumber\\
&\sum_{k=0}^m\sum_{i=1}^{n_k}p_{k,i}=1,
\nonumber\\
&\sum_{k=0}^m\sum_{i=1}^{n_k}
\exp\{\alpha_r+\bm q(y_{k,i})^\top\btheta_r\}p_{k,i}=1,
\quad r=1,\ldots,m,
\nonumber\\
&p_{k,i}>0,
\quad k=0,\ldots,m,\quad i=1,\ldots,n_k.
\nonumber
\end{align}

Direct maximization of the log-likelihood function \eqref{loglike0} is challenging because the unknown functions $\bm\beta$ are infinite-dimensional. To make the optimization tractable, we approximate each functional coefficient using a finite-dimensional basis expansion and estimate the corresponding basis coefficients. In this paper, we use B-spline basis functions because they provide a flexible representation for smooth functions and are straightforward to combine with roughness penalties. The resulting penalized B-spline estimators are introduced in the next subsection.

\subsection{Penalized B-spline estimator for $\beta_k(t)$}
For nonparametric estimation, the target function is typically assumed to be sufficiently smooth so that it can be well approximated by a finite basis expansion. B-splines are a common and effective choice for this purpose. For model \eqref{FPROP}, we use B-splines to approximate all $m+1$ unknown functions $\beta_0(t),\ldots,\beta_m(t)$:
\be
\beta_k(t) \approx \sum_{j=1}^{n_b}c_{k,j}b_j(t) = \bm b(t)^\top\bm c_k,\quad k=0, \ldots, m
\label{bspline4}
\ee
where $\bm b(t) = \left(b_1(t), \ldots, b_{n_b}(t)\right)^\top$ are B-spline basis functions and $\bm c_k = \left(c_{k,1}, \ldots, c_{k,n_b}\right)^\top$ are unknown coefficients. With the approximation \eqref{bspline4},
the inner product between $\beta_k(t)$ and $x_{k,i}(t)$ can be approximately expressed as
\[
\int \beta_k(t) x_{k,i}(t) dt \approx \sum_{j=1}^{n_b} c_{k,j} \int b_j(t) x_{k,i}(t) dt:= \bm u_{k,i}^\top\bm c_k,
\]
where $\bm u_{k,i} = \left(\int b_1(t) x_{k,i}(t) dt,\ldots, \int b_{n_b}(t) x_{k,i}(t) dt \right)^\top$.
Then the log-likelihood function defined in \eqref{loglike0}
can be approximately rewritten as:
\begin{align}
l_n(\bgamma,\bm c,\bm\alpha, \btheta,\bm p)
= & \sum_{k=0}^m\sum_{i=1}^{n_k}\left\{\bm z_{k,i}^\top\bgamma_k + \bm u_{k,i}^\top\bm c_k\right\} y_{k,i} + \sum_{k=0}^m\sum_{i=1}^{n_k}\log p_{k,i}  +\sum_{k=1}^m\sum_{i=1}^{n_k}\bm q(y_{k,i})^\top\btheta_k \nonumber\\
& - \sum_{k=0}^m\sum_{i=1}^{n_k} \log\left(\sum_{r=0}^m\sum_{j=1}^{n_r}p_{r,j}\text{exp}\left\{\left(\bm z_{k,i}^\top\bgamma_k + \bm u_{k,i}^\top\bm c_k\right) y_{r,j}  + \bm q(y_{r,j})^\top\btheta_k\right\}\right),\label{loglike1}\\
&s.t.\nonumber\\
&\sum_{k=0}^m\sum_{i=1}^{n_k} p_{k,i} = 1,\nonumber\\
&\sum_{k=0}^m\sum_{i=1}^{n_k} \exp\{\alpha_r + \bm q(y_{k,i})^\top\btheta_r\}p_{k,i} = 1,\quad r=1,\ldots,m,\nonumber\\
& p_{k,i}>0, \quad k=0,\ldots,m, \quad i=1,\ldots, n_k.\nonumber
\end{align}

To obtain an accurate B-spline approximation, the number of basis functions $n_b$ is often chosen to be moderately large. A large basis, however, can lead to unstable estimates and an ill-conditioned design matrix. We therefore regularize the coefficient vectors $\bm c_k$ using a roughness penalty, which controls the smoothness of the estimated functional coefficients and improves numerical stability.

In functional data analysis, a widely used choice is the integrated squared second-derivative penalty \citep{ramsayfunctional}, also commonly referred to as a roughness penalty:
\[
\text{Pen}\left\{\bm b(t)^\top\bm c_k\right\} = \|\left\{\bm b(t)^\top\bm c_k\right\}^{(2)}\|_2^2: = \bm c_k^\top \bm Q_k \bm c_k,
\]
where $\bm Q_k$ is a symmetric and positive semi-definite matrix.
This penalty discourages excessive curvature and favors smoother estimates. We therefore estimate the parameters by minimizing the following penalized negative empirical log-likelihood with respect to $\left(\bgamma,\bm c,\bm\alpha,\btheta,\bm p\right)$:
\begin{align}
\min& -l_n(\bgamma,\bm c,\bm\alpha, \btheta,\bm p)  + \sum_{k=0}^m \lambda_k \bm c_k^\top\bm Q_k\bm c_k,\label{penloglike}\\
&s.t.\nonumber\\
&\sum_{k=0}^m\sum_{i=1}^{n_k} p_{k,i} = 1,\nonumber\\
&\sum_{k=0}^m\sum_{i=1}^{n_k} \exp\{\alpha_r + \bm q(y_{k,i})^\top\btheta_r\}p_{k,i} = 1, \quad r=1,\ldots,m,\nonumber\\
& p_{k,i}>0, \quad k=0,\ldots,m,\quad i=1,\ldots, n_k,\nonumber
\end{align}
where $\lambda_k\geq 0$ are smoothing parameters. The penalized criterion balances goodness of fit against smoothness of the estimated functional coefficients.

\subsection{Choice of the density ratio basis}
The vector $\bm q(y)$ specifies the structure of the density ratios between each population-specific baseline distribution and the reference baseline distribution. A simple specification of $\bm q(y)$ is attractive because it introduces relatively few nuisance parameters, but it may be too restrictive when the baseline distributions differ in more complex ways. Conversely, an unnecessarily rich specification increases the dimension of $\btheta_k$ and may lead to less stable estimation, particularly when the population-specific sample sizes are small. In practice, the choice of $\bm q(y)$ should therefore balance flexibility and parsimony.

Several commonly used parametric distribution families motivate useful choices of $\bm q(y)$. For example, $(y,y^2)^\top$ is sufficient to represent density ratios between Normal distributions with different means and variances, while combinations involving $y$ and $\log y$ arise naturally for positive-valued distributions such as Gamma distributions. More flexible polynomial or transformed bases can be used when the shape of the baseline distributions is less certain. Because the normalizing constants $\alpha_k$ already play the role of intercepts in the density-ratio representation, a constant term does not need to be included in $\bm q(y)$.

In applications, one possible strategy is to compare a small collection of scientifically or distributionally motivated candidate bases using an information criterion or a sensitivity analysis. The main inferential conclusions should preferably be stable over reasonable choices of $\bm q(y)$. In the simulation studies below, we use the basis corresponding to the true data-generating density ratio so that the simulations focus on the efficiency and robustness of the proposed estimation procedure rather than on misspecification of $\bm q(y)$. In the soybean application, where the baseline distribution is unknown, we use a richer basis and examine the sensitivity of the results to alternative specifications.

\subsection{Computational implementation}
After the B-spline approximation in \eqref{bspline4}, the infinite-dimensional estimation problem is reduced to a finite-dimensional constrained optimization problem in $(\bgamma,\bm c,\bm\alpha,\btheta,\bm p)$. For fixed smoothing parameters $(\lambda_0,\ldots,\lambda_m)$, all observed responses from the $m+1$ populations are pooled to form the support of the empirical baseline distribution $G_0$. The constraints in \eqref{penloglike} ensure that the masses $\bm p$ define a probability distribution and that each tilted baseline distribution $G_k$ is properly normalized.

Once $(\hat{\bm\alpha},\hat\btheta,\hat{\bm p})$ has been obtained, the estimated probability mass assigned by population $k$ to a pooled response value $y_{r,j}$ is $\hat p_{r,j}\exp\{\hat\alpha_k+\bm q(y_{r,j})^\top\hat\btheta_k\}$,
with $\alpha_0=0$ and $\btheta_0=\bm 0$. Hence, observations from one population can contribute to estimation of the baseline distribution for another population through the DRM weights. This pooled-support construction is the main computational mechanism through which the proposed method borrows distributional information across populations.

For a given set of smoothing parameters, we numerically minimize the penalized negative empirical log-likelihood in \eqref{penloglike} subject to the normalization and positivity constraints. The resulting estimates are then used in the cross-validation procedure described in the next subsection.

\subsection{Parameter tuning strategy for $\lambda_k$}
We use cross-validation to select the smoothing parameters in model \eqref{penloglike}. Let $I_k$ and $I_k^*$ denote the training and test index sets, respectively, for population $k$.
Define 
\[
S_k=\{(\bm z_{k,i}, x_{k,i}(t), y_{k,i}):i\in I_k\}, \quad k=0, \ldots, m,
\]
and 
\[
S_k^*=\{(\bm z_{k,i}, x_{k,i}(t), y_{k,i}):i\in I_k^*\}, \quad k=0, \ldots, m.
\]
To maintain the ratio of sample sizes from different populations in the cross-validation procedure, 
we ensure that
\[
\frac{|I_k|}{|I_0|} = \frac{n_k}{n_0}, \quad k=1, \ldots, m,
\]
when constructing the training sets $S_k$.

For each validation fold, we fit model \eqref{penloglike} using the pooled training set $\cup_{k=0}^m S_k$ to obtain $(\hat\bgamma,\hat{\bm c},\hat{\bm\alpha},\hat{\btheta},\hat{\bm p})$. We then evaluate the negative log-likelihood on the pooled test set $\cup_{k=0}^m S_k^*$ and use it as the validation loss. The smoothing parameters $(\lambda_0,\ldots,\lambda_m)$ are selected by 10-fold cross-validation to minimize the average validation loss.

To calculate the log-likelihood of $S_k^*$, we need to compute $P_{\hat G_k(y)}(Y=y_{k,i})$ for $i\in I_k^*$, $k=0, \ldots, m$, based on $\hat G_k(y)$ estimated from $\cup_{k=0}^m S_k$.
Because the empirical likelihood estimator is discrete and supported only on the training responses, a test response $y_{k,i}$ that is not observed in the training data would receive zero probability mass. Direct use of this mass in the test likelihood would therefore be inappropriate. To evaluate out-of-sample likelihoods, we first construct a continuous approximation to the estimated baseline density using kernel smoothing and then use it to approximate the probability mass associated with each test response.
Define a kernel smoothed EL estimator for the density function $g_k(y)$ of $G_k(y)$ as
\be
\hat g_k^{ksel}(y) = \frac{1}{h}\sum_{r=0}^m\sum_{i\in I_r}\hat p_{r,i}\text{exp}\left\{\hat\alpha_k + \bm q(y_{r,i})^\top\hat\btheta_k\right\}K\left(\frac{y-y_{r,i}}{h}\right),
\label{ksel}
\ee
where $K(\cdot)$ is a kernel function and $h$ is the bandwidth.

Next, we use a local weighted average of the probability masses assigned to training observations near the target test response $y_{k,i}$. Specifically, for any $i\in I_k^*$, define the neighborhood
\[
N_{\eta}(y_{k,i}):= \{y_{r,j}: |y_{r,j}-y_{k,i}|\leq \eta, \quad j\in I_r, \quad r=0,\ldots, m\}.
\]
Then we estimate $p_{k,i}$ for $i\in I_k^*$ by
\be
\tilde p_{k,i} = \frac{1}{|N_{\eta}(y_{k,i})|}\sum_{y_{r,j}\in N_{\eta}(y_{k,i})}\frac{\hat g_k^{ksel}(y_{k,i})}{\hat g_k^{ksel}(y_{r,j})}\hat p_{r,j}\text{exp}\left\{\hat\alpha_k + \bm q(y_{r,j})^\top\hat\btheta_k\right\}, \quad k=0,\ldots, m. \label{cvtune}
\ee
This construction provides a nonzero local approximation to $p_{k,i}$ for $i\in I_k^*$, $k=0,\ldots,m$, allowing the empirical likelihood contributions of the test observations to be evaluated using information from the training sample.

We implement 10-fold cross-validation as follows:
\begin{itemize}
\item[] 1. Partition the observations within each population into 10 folds while approximately preserving the relative sample sizes across populations.

\item[] 2. For a candidate combination $(\lambda_0,\ldots,\lambda_m)$, estimate $(\bgamma,\bm c,\bm\alpha,\btheta,\bm p)$ by maximum penalized empirical likelihood using the pooled training data $\cup_{k=0}^m S_k$.

\item[] 3. For each observation in the pooled test data $\cup_{k=0}^m S_k^*$, use \eqref{ksel} and \eqref{cvtune} to obtain the smoothed approximation $\tilde p_{k,i}$ and compute the corresponding negative log-likelihood.

\item[] 4. Repeat Steps 1--3 over all 10 folds and average the validation loss.

\item[] 5. Select the combination $(\lambda_0,\ldots,\lambda_m)$ that yields the smallest average validation loss, and refit the model to the full data using the selected smoothing parameters.
\end{itemize}

\section{Theoretical Results}

Without loss of generality, we establish consistency with one common smoothing parameter
$\lambda=\lambda_N$ for all roughness penalties. Let $N=\sum_{k=0}^m n_k$, with
$m$ fixed, and write the true parameter as
$\mathcal P^0=(\bgamma^0,\bm\beta^0,\bm\alpha^0,\btheta^0,G_0^0)$.
The functional estimates are
$\hat\beta_k(t)=\bm b(t)^\top\hat{\bm c}_k$. The baseline estimates are
\[
\hat G_0(y)=\sum_{r=0}^m\sum_{j=1}^{n_r}
\hat p_{r,j}\,\mathbbm{1}(y_{r,j}\le y),
\]
and, for $k=1,\ldots,m$,
\[
\hat G_k(y)=\sum_{r=0}^m\sum_{j=1}^{n_r}
\hat p_{r,j}e^{\hat\alpha_k+\bm q(y_{r,j})^\top\hat\btheta_k}
\mathbbm{1}(y_{r,j}\le y).
\]
The DRM normalization determines
\[
\hat\alpha_k=-\log\left\{
\sum_{r=0}^m\sum_{j=1}^{n_r}
\hat p_{r,j}e^{\bm q(y_{r,j})^\top\hat\btheta_k}
\right\},\qquad k=1,\ldots,m,
\]
with $\hat\alpha_0=0$ and $\hat\btheta_0=0$.
Let $\Xi$ be a compact set for
$(\bgamma,\btheta)$ containing $(\bgamma^0,\btheta^0)$ in its interior,
and let
\[
\mathcal B_N=
\left\{\beta\in\operatorname{span}\{b_1,\ldots,b_{n_b}\}:
\|\beta\|_{L^2}\le R,\quad
\|\beta^{(2)}\|_{L^2}^2\le C\right\},
\]
where the fixed constants satisfy
\[
R>\max_{0\le k\le m}\|\beta_k^0\|_{L^2},\qquad
C>\max_{0\le k\le m}\|(\beta_k^0)^{(2)}\|_{L^2}^2.
\]
Let $V:\mathbb R\to[1,\infty)$ be a continuous weight function
satisfying $V(y)\to\infty$ as $|y|\to\infty$.
Choose a fixed constant $M_G$ such that $\int V(y)\,dG_0^0(y)<M_G<\infty$.
In addition to the positivity and probability normalization constraints
for the baseline distributions, we impose the following restrictions:
\begin{equation}
\begin{aligned}
& (\bgamma,\btheta)\in\Xi,\qquad
\beta_k\in\mathcal B_N\quad(k=0,\ldots,m),\\
& \sum_{r=0}^m\sum_{j=1}^{n_r}p_{r,j}V(y_{r,j})\le M_G.
\end{aligned}
\label{eq:theory-restrictions}
\end{equation}
Denote this empirical admissible parameter space by $\mathcal A_N$.
Conditions (C1)--(C6) in the Supplementary Materials specify the sampling,
identification, tail-integrability, spline-feasibility, and smoothing
requirements.

\begin{theorem}[Consistency]\label{thm:consistency}
Suppose Conditions (C1)--(C6) in the Supplementary Material hold.
In particular, $n_b\to\infty$, $\lambda_N\ge0$, and
$\lambda_N/N\xrightarrow{p}0$. Let $\mathcal Q_N$ denote the
unnormalized penalized objective in \eqref{penloglike}, with the common
smoothing parameter $\lambda_N$. Let
\[
\hat{\bm\vartheta}_N=
(\hat\bgamma,\hat{\bm c},\hat{\bm\alpha},\hat\btheta,\hat{\bm p})
\in\mathcal A_N
\]
be an exact minimizer or a feasible approximate minimizer satisfying
\[
\mathcal Q_N(\hat{\bm\vartheta}_N)
\le\inf_{\bm\vartheta\in\mathcal A_N}\mathcal Q_N(\bm\vartheta)+e_N,
\qquad e_N\ge0,\quad e_N=o_p(N).
\]
Then
\[
\begin{aligned}
&\|\hat\bgamma-\bgamma^0\|+
\|\hat{\bm\alpha}-\bm\alpha^0\|+
\|\hat\btheta-\btheta^0\|\\
&\qquad+\max_{0\le k\le m}
\|\hat\beta_k-\beta_k^0\|_{L^2}\xrightarrow{p}0.
\end{aligned}
\]
Moreover, $\hat G_k\Rightarrow G_k^0$ in probability for
$k=0,\ldots,m$. If $G_0^0$ is continuous, then all true DRM baselines
are continuous and
\[
\max_{0\le k\le m}\sup_y
|\hat G_k(y)-G_k^0(y)|\xrightarrow{p}0.
\]
\end{theorem}

The proof is provided in the Supplementary Material.

\section{Simulation Studies}
\subsection{Data Generating Models}
We conduct three simulation studies to evaluate different aspects of the proposed method. Scenario I is designed to isolate the efficiency gain from borrowing information across related populations by comparing the proposed DRM-EL estimator with separate empirical likelihood fitting. Scenario II considers a Gaussian response model and compares the proposed semiparametric approach with a conventional Gaussian functional regression fit when the parametric response model is correctly specified. Scenario III replaces the Gaussian response with a truncated Normal distribution and is used to examine robustness when the Gaussian response assumption used by the competing parametric method is misspecified. Together, the three settings distinguish the benefit of multi-population borrowing from the benefit of avoiding a restrictive parametric response distribution.

In all three scenarios, the functional predictor is generated from a Wiener process on $[0,1]$. This produces stochastic functional observations with nontrivial within-curve dependence and allows the performance of the functional coefficient estimators to be evaluated without reducing the predictor to a finite-dimensional construction at the data-generating stage. The regression functions and scalar coefficients vary across populations in selected scenarios so that the populations remain related but need not have identical predictor--response relationships.

\noindent\textbf{Scenario I.}
In the first scenario, the data sets are generated from the following model:
\be
Y_{k,i} = \int_{[0,1]}\rho_k(t)X_{k,i}(t)dt + \epsilon_{k,i}, \quad k=0,1,
\label{sim1}
\ee
where $X_{k,i}(t)$ is generated from a Wiener process, $\epsilon_{k,i}\sim N(0,\sigma_k^2)$ with $\sigma_0=0.1$ and $\sigma_1=0.09$, and
\[
\rho_0(t) = 10(t-0.5)^2,\quad \rho_1(t) = 8(t-0.4)^2.
\]

\noindent\textbf{Scenario II.}
In the second scenario, the data sets are generated from the following model:
\be
Y_{k,i} = \nu_{k,1}Z_{1,k,i} + \nu_{k,2}Z_{2,k,i} + \int_{[0,1]}\rho_k(t)X_{k,i}(t)dt + \epsilon_{k,i}, \quad k=0,1,2
\label{sim2}
\ee
where $X_{k,i}(t)$ is generated from a Wiener process; for $k=0, 1, 2$, $Z_{1,k,i} \sim Unif(-0.5, 0.5)$, $Z_{2,k,i} \sim Unif(-0.5, 0.5)$, $\epsilon_{k,i}\sim N(\mu_k,\sigma_k^2)$ with $(\mu_0, \mu_1, \mu_2) = (0, 0, 0.05)$ and $(\sigma_0, \sigma_1, \sigma_2 ) = (0.1, 0.1, 0.2)$; and
\[
\bm\nu_0 = (0.2, 0.4)^\top,\quad \bm\nu_1 = (0.4, 0.2)^\top, \quad \bm\nu_2 = (0.2, 0.4)^\top;
\]
\[
\rho_k(t) = 10(t-0.5)^2, \quad k=0, 1, 2.
\]

\noindent\textbf{Scenario III.}
In the third scenario, the data sets are generated from the following model:
\be
Y_{k,i}|\bm Z_{k,i}, X_{k,i}(t)\sim TN(\mu_{k,i},\sigma_k), \quad \mu_{k,i} = \bm\nu_k^\top\bm Z_{k, i} + \int_0^1\rho_k(t)X_{k,i}(t)dt, \quad k=0,1,2
\label{sim3}
\ee
where $TN(\mu_{k,i},\sigma_k)$ denotes a truncated Normal r.v. over $[0, \infty)$ with $(\sigma_0, \sigma_1, \sigma_2 ) = (0.1, 0.1, 0.15)$; $X_{k,i}(t)$ is generated from a Wiener process; $Z_{1,k,i} \sim Unif(-0.5, 0.5)$, $Z_{2,k,i} \sim Unif(-0.5, 0.5)$;
\[
\bm\nu_0 = (0.2, 0.4)^\top,\quad \bm\nu_1 = (0.4, 0.2)^\top, \quad \bm\nu_2 = (0.1, 0.1)^\top,
\]
\[
\rho_k(t) = 10(t-0.5)^2,\quad k=0, 1, 2.
\]

\subsection{Simulation Results}
Under data-generating model \eqref{sim1}, the conditional distribution of $Y_{k,i}$ is Normal. Therefore, model \eqref{sim1} can be expressed in the form of the proposed model \eqref{fprop_drm}, with
% \[
% f_k(y|x_{k,i}(t)) = \frac{\text{exp}\{\langle x_{k,i},\rho_k(t)\rangle/\sigma_k^2\}yf_{\epsilon,k}(y)}{\int_0^1 \text{exp}\{\langle x_{k,i},\rho_k\rangle/\sigma_k^2\}yf_{\epsilon,k}(y)}
% =\frac{\text{exp}\{\langle x_{k,i},\beta_k\rangle y_{k,i}\}f_{\epsilon,k}(y_{k,i})}{\int_0^1\text{exp}\{\langle x_{k,i},\beta_k\rangle y_{k,i}\}f_{\epsilon,k}(y_{k,i})},
% \]
% where $f_{\epsilon,k}(y)$ is the density function of $\epsilon_{k,i}$, and
\[
\beta_0(t)=\rho_0(t)/\sigma_0^2, \quad \beta_1(t)=\rho_1(t)/\sigma_1^2.
\]
We consider two sample-size settings for model \eqref{sim1}: $(n_0,n_1)=(100,200)$ and $(n_0,n_1)=(200,200)$. Each setting is repeated 100 times. Because the $L_2$-norms of the true functions $\beta_0(t)$ and $\beta_1(t)$ are relatively large, we use the relative integrated squared error (RISE) to evaluate the penalized B-spline estimators:
\[
\text{RISE}:=\frac{\|\hat\beta_k(t)-\beta_k(t)\|^2_{L2}}{\|\beta_k(t)\|^2_{L2}}.
\]
Under the proposed method, we utilize the true basis function $\bm q(y) = (y, y^2)^\top$ for the DRM framework. In the cross-validation procedure for tuning parameters, we employ
the Epanechnikov kernel,
\[
K(u)=\frac{3}{4}(1-u^2)\mathbbm{1}\{|u|\leq 1\},
\]
for the kernel-smoothed EL density estimator, denoted by
$\hat g_k^{ksel}(y)$. Within each training fold, we select a population-specific bandwidth using the following rule \citep{silverman2018density},
\[
h_k
=0.9\,|I_k|^{-1/5}
\min\left\{
\hat\sigma_k,\,
\frac{\widehat{\mathrm{IQR}}_k}{1.34}
\right\},
\qquad k=0,\ldots,m,
\]
where $|I_k|$ is the training sample size for population $k$, and
$\hat\sigma_k$ and $\widehat{\mathrm{IQR}}_k$ are the standard
deviation and interquartile range of the fitted baseline
distribution $\hat G_k$, respectively. All quantities used to
select the bandwidth are calculated from the training data only.

\begin{table}[!htbp]
\centering
\renewcommand\arraystretch{1}
\resizebox{\linewidth}{!}{%
\begin{tabular}{rccccccc}
\toprule
Methods & $(n_0, n_1)$ & RISE($\hat\beta_0$) & RISE($\hat\beta_1$) & SE($\hat\theta_{1,1}$) & Bias($\hat\theta_{1,1}$) & SE($\hat\theta_{1,2}$) & Bias($\hat\theta_{1,2}$)\\
\hline
EL    &  $(100, 200)$ & 0.263 & 0.129 & & & &\\
DRM-EL &  $(100, 200)$ & 0.224 & 0.104 & 1.447 & 0.035 & 11.915& 2.068\\
EL    &  $(200, 200)$ & 0.101 & 0.129 & & & &\\
DRM-EL &  $(200, 200)$ & 0.059 & 0.054 & 1.171 & 0.034 & 9.058& -1.513\\
\bottomrule
\end{tabular}%
}
\caption{Simulation results based on Scenario I for the two proposed methods: EL and DRM-EL.}
\label{table:sim1}
\end{table}

For Scenario I, we compare two methods: EL, which fits model \eqref{fprop} separately to each population using maximum empirical likelihood, and DRM-EL, which fits the proposed multi-population model \eqref{fprop_drm} using the pooled samples. Thus, EL corresponds to fitting the semiparametric model without borrowing information across populations, whereas DRM-EL uses the density ratio structure to connect the baseline distributions.

\begin{figure}[!htbp]
  \centering
   \begin{subfigure}{0.45\textwidth}
   \centering
   \includegraphics[width=\linewidth]{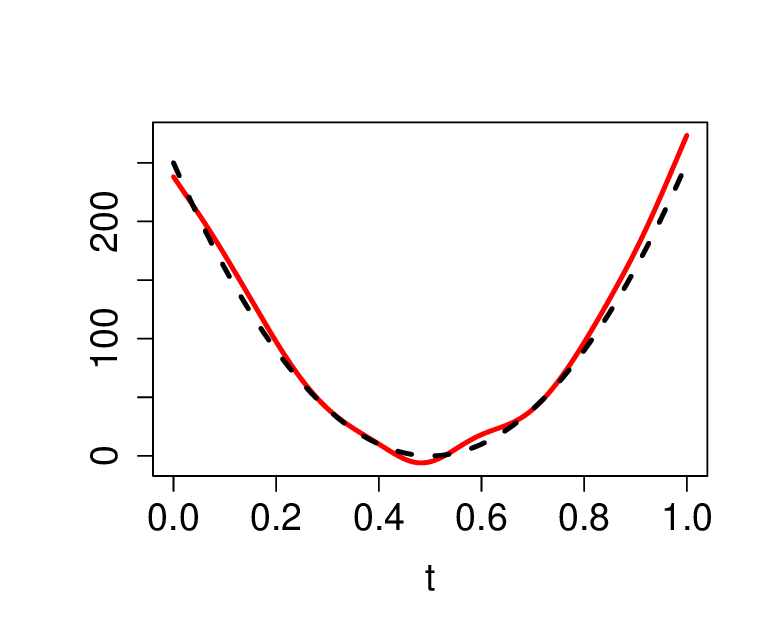}
   \caption{$\hat\beta_0(t)$ obtained from EL method}
   \end{subfigure}
   \begin{subfigure}{0.45\textwidth}
   \centering
   \includegraphics[width=\linewidth]{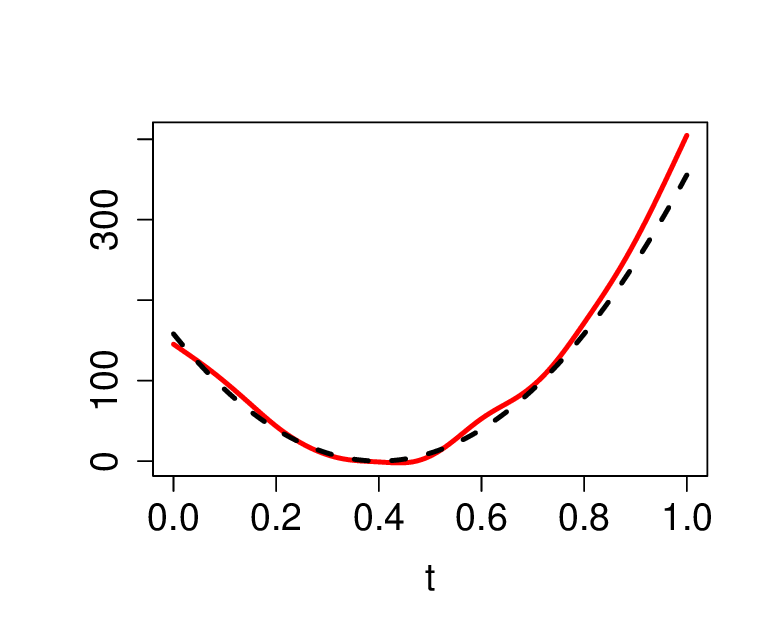}
   \caption{$\hat\beta_1(t)$ obtained from EL method}
   \end{subfigure}
   \newline
   \begin{subfigure}{0.45\textwidth}
   \centering
   \includegraphics[width=\linewidth]{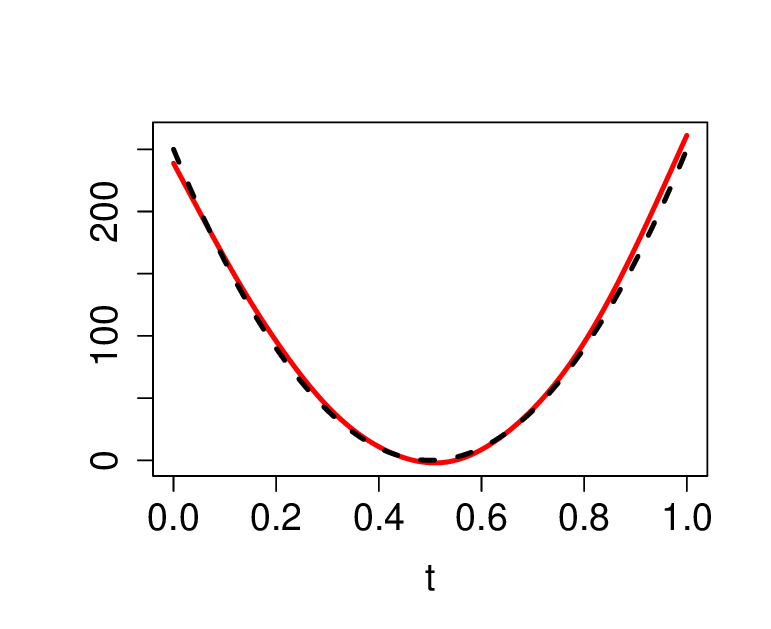}
   \caption{$\hat\beta_0(t)$ obtained from DRM-EL method}
   \end{subfigure}
   \begin{subfigure}{0.45\textwidth}
   \centering
   \includegraphics[width=\linewidth]{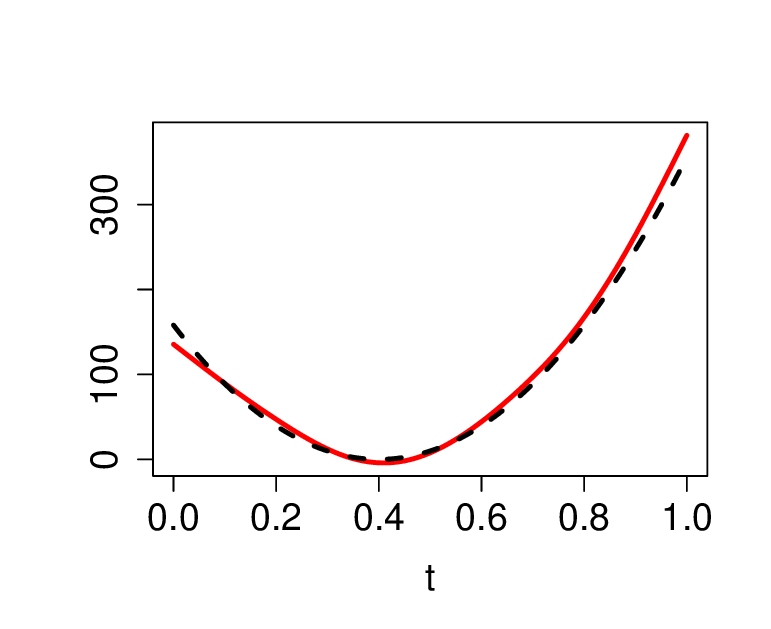}
   \caption{$\hat\beta_1(t)$ obtained from DRM-EL method}
   \end{subfigure}
  \caption{Averaged $\hat\beta_0(t)$ and $\hat\beta_1(t)$ (red solid line) obtained from two proposed methods: EL and DRM-EL based on 100 repetitions under the setting $(n_0, n_1) = (200, 200)$; the true $\beta_0(t)$ and $\beta_1(t)$ (black dashed line).}
  \label{fig:fglm-el-sim1}
\end{figure}

Table \ref{table:sim1} shows that DRM-EL produces smaller RISE values than separate EL fitting for both functional coefficients under both sample-size settings. For $(n_0,n_1)=(100,200)$, the RISE decreases from 0.263 to 0.224 for $\beta_0(t)$ and from 0.129 to 0.104 for $\beta_1(t)$, corresponding to reductions of approximately 15\% and 19\%, respectively. The improvement is more pronounced when both populations have $n_k=200$: the RISE decreases from 0.101 to 0.059 for $\beta_0(t)$ and from 0.129 to 0.054 for $\beta_1(t)$, reductions of approximately 42\% and 58\%, respectively. These results indicate that the density ratio structure can improve estimation efficiency by borrowing information across related populations while retaining population-specific regression functions. 
Figure \ref{fig:fglm-el-sim1} presents the estimated coefficient functions $\beta_0(t)$ and $\beta_1(t)$ averaged over 100 repetitions under the setting $(n_0,n_1)=(200,200)$. The DRM-EL estimates follow the true coefficient functions more closely than the separate EL estimates, particularly for $\beta_0(t)$, suggesting that borrowing information through the density ratio structure reduces estimation bias.
It is worth noting that Scenario I uses the correctly specified DRM basis $\bm q(y)=(y,y^2)^\top$. Thus, this comparison is intended primarily to assess the potential efficiency gain from joint estimation when the density-ratio relationship is adequately represented, rather than robustness to misspecification of $\bm q(y)$. Sensitivity to the choice of the density-ratio basis is instead considered in the real-data analysis, where the true baseline distributions are unknown.

For Scenarios II and III, we use the same sample-size configuration, $(n_0,n_1,n_2)=(100,50,50)$, and conduct 100 repetitions for each scenario. Both data-generating models \eqref{sim2} and \eqref{sim3} can be represented in the form of \eqref{fprop_drm}, with
$\beta_k(t)=\rho_k(t)/\sigma_k^2$, $\bgamma_k = \bm\nu_k/\sigma_k^2$, for $k=0, 1, 2$,  where $\bm\nu_k=(\nu_{k,1},\nu_{k,2})^\top$.
The primary objective of Scenarios II and III is to compare the estimation accuracy of the functional coefficients $\beta_k(t)$ and scalar coefficients $\bgamma_k$ obtained using the proposed DRM-EL method and Gaussian maximum likelihood estimation (MLE). The Gaussian response distribution is correctly specified in Scenario II and misspecified in Scenario III.
For the competing Gaussian MLE method, we use the \texttt{fregre.glm} function from the R package \texttt{fda.usc}. The functional coefficients are approximated using B-spline basis expansions, with the number of basis functions selected by the Akaike information criterion (AIC). In this implementation, no additional roughness penalty is imposed on the functional coefficients.

Tables~\ref{table:sim23-1} and~\ref{table:sim23-2} summarize the simulation results. In both scenarios, DRM-EL yields substantially smaller RISE values for the functional coefficients than Gaussian MLE. This difference may partly reflect the different approaches to controlling smoothness: DRM-EL incorporates a roughness penalty, whereas the Gaussian MLE implementation relies on AIC-based selection of the basis dimension. Consequently, these results compare the two estimation procedures as implemented, and the improvement cannot be attributed solely to the density ratio structure.

For the scalar coefficients, the results under Scenario II are mixed: Gaussian MLE has smaller MSEs for three of the six coefficients, while DRM-EL has smaller MSEs for the remaining three. Thus, DRM-EL remains competitive when the Gaussian distribution is correctly specified, potentially benefiting from information sharing across populations through the DRM structure in \eqref{fprop_drm}. Under Scenario III, DRM-EL yields smaller MSEs for all six scalar coefficients. Together with the results for the functional coefficients, these findings demonstrate the robustness of the proposed semiparametric approach in the misspecified setting considered here.

\begin{table}[htbp]
\centering
\renewcommand\arraystretch{1}
\begin{tabular}{lcccc}
\toprule
Scenario & Method & RISE($\hat\beta_0$) & RISE($\hat\beta_1$) & RISE($\hat\beta_2$)\\
\midrule
\multirow{2}{*}{II} & DRM-EL & 0.12 & 0.13 & 0.16\\
 & MLE & 3.05 & 5.78 & 10.29\\
\midrule
\multirow{2}{*}{III} & DRM-EL & 0.20 & 0.22 & 0.47\\
 & MLE & 4.59 & 14.01 & 16.05\\
\bottomrule
\end{tabular}
\caption{RISE values for the estimated functional coefficients under Scenarios II and III. DRM-EL denotes the proposed method based on model \eqref{fprop_drm}, and Gaussian MLE denotes maximum likelihood estimation assuming a Gaussian response distribution.}
\label{table:sim23-1}
\end{table}

\begin{table}[htbp]
\centering
\renewcommand\arraystretch{1}
\begin{tabular}{lcccc}
\toprule
Scenario & Method
& $\bigl\{\mathrm{MSE}(\hat\gamma_{0,1}),\mathrm{MSE}(\hat\gamma_{0,2})\bigr\}$
& $\bigl\{\mathrm{MSE}(\hat\gamma_{1,1}),\mathrm{MSE}(\hat\gamma_{1,2})\bigr\}$
& $\bigl\{\mathrm{MSE}(\hat\gamma_{2,1}),\mathrm{MSE}(\hat\gamma_{2,2})\bigr\}$\\
\midrule
\multirow{2}{*}{II} & DRM-EL
& $(0.025, 0.015)$ & $(0.011, 0.027)$ & $(0.037, 0.076)$\\
 & MLE
& $(0.018, 0.022)$ & $(0.022, 0.022)$ & $(0.042, 0.025)$\\
\midrule
\multirow{2}{*}{III} & DRM-EL
& $(0.020, 0.027)$ & $(0.027, 0.030)$ & $(0.011, 0.032)$\\
 & MLE
& $(0.075, 0.031)$ & $(0.056, 0.070)$ & $(0.030, 0.044)$\\
\bottomrule
\end{tabular}
\caption{Componentwise MSEs for the estimated scalar coefficients under Scenarios II and III. Each pair reports the MSEs for the two components of $\hat\bgamma_k$. DRM-EL denotes the proposed method based on model \eqref{fprop_drm}, and MLE denotes maximum likelihood estimation assuming a Gaussian response distribution.}
\label{table:sim23-2}
\end{table}

\section{Real Data Analysis}
Kansas is a major soybean-producing state in the United States, where growing-season environmental conditions play an important role in crop production. Previous studies have highlighted the effects of temperature and water availability on soybean growth and yield \citep{vogel2019effects, schlenker2009nonlinear}. Although soybeans can tolerate moderate drought stress during vegetative growth, heat and water stress during reproductive stages can increase flower and pod abortion and reduce final yield. Low temperatures can also impair germination and growth. These considerations motivate an analysis of soybean yield that jointly accounts for temperature trajectories and irrigation.

\subsection{Data description and preprocessing}

We apply the proposed method to county-level soybean data from Kansas. Annual soybean yield and agricultural information for 1991--2006 were obtained from the United States Department of Agriculture (USDA) Quick Stats database (\url{https://quickstats.nass.usda.gov/}). These data include soybean yield, measured in bushels per acre, harvested acreage, and irrigated acreage. Corresponding climate data were obtained from the National Oceanic and Atmospheric Administration (NOAA; \url{https://www.ncei.noaa.gov/products}) and include daily minimum and maximum temperatures at individual weather stations.

To integrate the agricultural and climate data, we assigned each weather station to the county whose center was geographically closest to the station, using the latitude and longitude of the station and county centers. We then averaged the station-level climate measurements within each assigned county to obtain county-level daily minimum and maximum temperatures. The USDA records cover a subset of Kansas counties in each year.

The response is annual soybean yield. We include the proportion of harvested acreage that is irrigated as one scalar predictor.
The functional predictor is the daily temperature trajectory, defined at each day as the average of the county-level daily minimum and maximum temperatures.

For the analysis, we select four years: 1993, 1998, 2001, and 2004. These years provide adequate sample sizes while spanning much of the period covered by the available data. We treat each year as a separate population to account for differences in annual environmental and management conditions. The populations nevertheless share the same crop, geographic setting, and variable definitions, making information sharing across years plausible. The proposed DRM framework accommodates differences in the baseline yield distributions while borrowing information across these related populations.
We define the variables as follows:
\bit
\item $X(t)$ represents the daily average temperature from June through November.
\item $Z$ represents the proportion of irrigated area for each county in Kansas.
\item $Y$ represents the annual soybean yield per unit area for each county in Kansas.
\eit

\subsection{Model specification and estimation}
We index the four years by $k=0,1,2,3$, corresponding to 1993, 1998, 2001, and 2004, respectively. We fit the following model:
\begin{align}
f_k(y|z, x(t))&=\frac{\text{exp}\left\{\left(\gamma_k z+ \langle x, \beta_k\rangle\right) y\right\}g_k(y)}{\int \text{exp}\left\{\left(\gamma_k z + \langle x, \beta_k\rangle\right) y\right\}g_k(y)dy},\quad k=0,1,2,3,\label{soybean_model4}\\
\frac{g_r(y)}{g_0(y)} &= \exp\{\alpha_r + \bm\theta_r^\top \bm q(y)\},\quad r=1,2,3.\nonumber
\end{align}
For the density ratio structure in \eqref{soybean_model4}, we use $\bm q(y) = (y, y^2, y^3, \log(y))^\top$,
which is sufficiently flexible to include terms associated with both Normal- and Gamma-type density ratios. Because the true form of the between-year baseline density ratios is unknown, we also considered richer specifications containing additional polynomial terms. The estimated functional-coefficient patterns and the qualitative conclusions for the irrigation effects were very similar, suggesting that the main findings are not driven by the particular basis reported above. We therefore present results based on $\bm q(y)=(y,y^2,y^3,\log y)^\top$ for parsimony.

\subsection{Results and interpretation}
\begin{figure}[!htbp]
  \centering
  \includegraphics[width=0.45\linewidth]{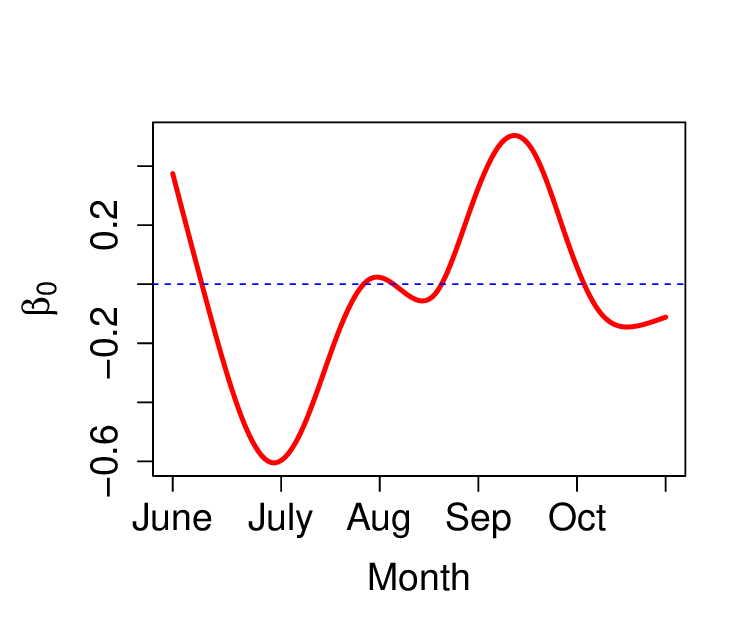}
  \includegraphics[width=0.45\linewidth]{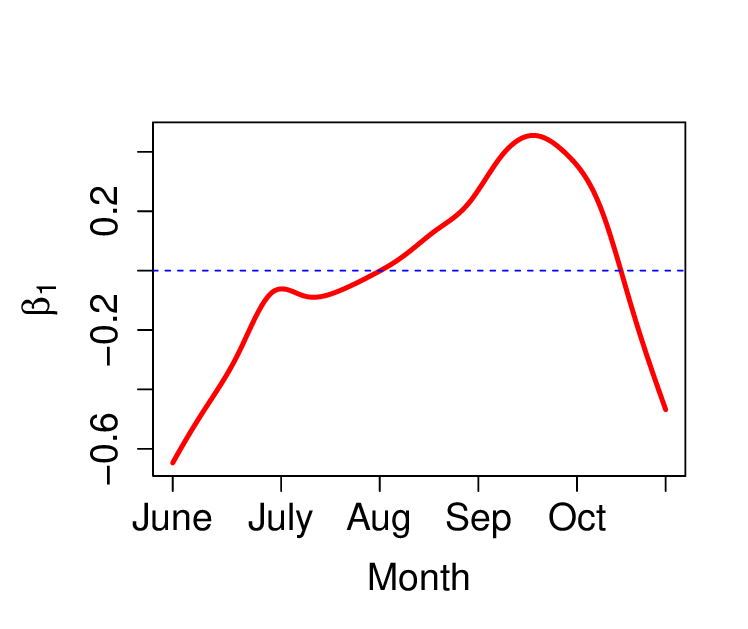}
  %\medskip
  \includegraphics[width=0.45\linewidth]{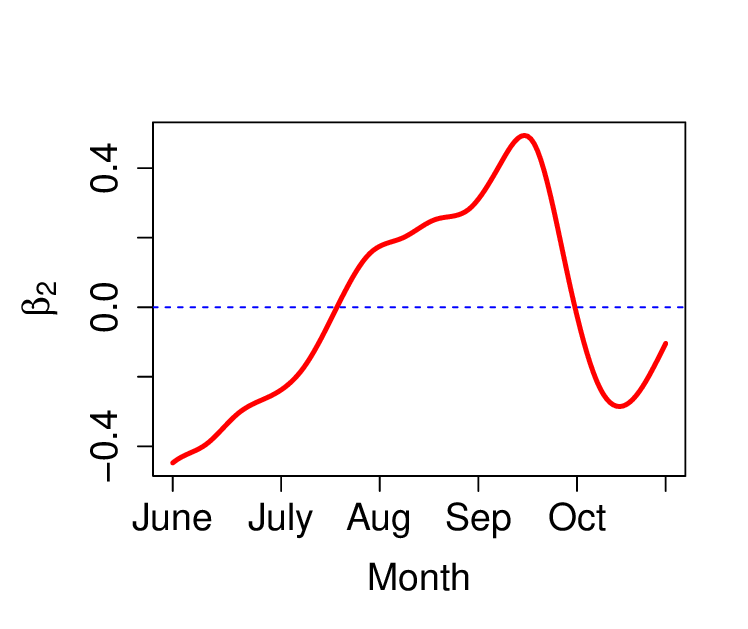}
  \includegraphics[width=0.45\linewidth]{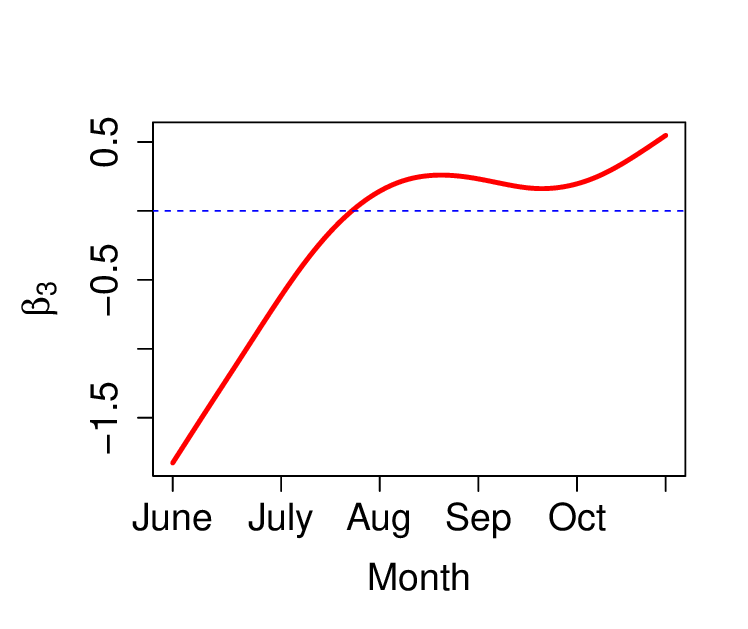}
  \caption{Estimated functional coefficients $\hat\beta_k(t)$ obtained using the proposed DRM-EL method for the Kansas soybean data.}
\label{fig:beta_soybean}
\end{figure}

The estimated functional coefficients $\hat\beta_k(t)$, $k=0,1,2,3$, are displayed in Figure \ref{fig:beta_soybean}. The four estimated curves exhibit a broadly consistent temporal pattern. 
The estimated functional coefficients are generally negative around July and positive during much of August–September, with transitions occurring around late July to early August and additional sign changes that vary across years. These patterns indicate that the direction of the association between temperature and soybean yield, after accounting for irrigation, varies over the growing season and across years. The negative associations around July may reflect unfavorable effects of higher temperatures during an already warm period, whereas the positive associations during much of August–September suggest that relatively warmer conditions may be beneficial as seasonal temperatures decline. These associations may also reflect differences in temperature sensitivity across soybean growth stages, including flowering, pod formation, and seed filling. However, because growth-stage information is unavailable, the observed temporal patterns cannot be attributed directly to specific developmental stages.

Under model \eqref{soybean_model4}, the pointwise value $\beta_k(t)$ is most naturally interpreted through its contribution to the exponential-tilting index. Holding the remainder of the temperature trajectory and the irrigation variable fixed, a higher temperature at time $t$ increases the index when $\beta_k(t)>0$ and decreases it when $\beta_k(t)<0$. As discussed in Section 1, an increase in the index shifts relative probability toward larger yield values. Therefore, the sign of $\hat\beta_k(t)$ provides a distributional interpretation of the direction of association at each time point, rather than representing a conventional additive change in mean yield per unit increase in temperature.

The similarity in the shapes of the four estimated curves suggests that the timing of the temperature--yield association is reasonably consistent across the selected years, even though the baseline response distributions and the magnitudes of the effects are allowed to differ. This illustrates an important feature of the proposed analysis: the model does not require the functional coefficients to be identical across years, but the estimated curves can still reveal common temporal features when such features are supported by the data.

The estimated coefficients for the irrigation variable are $(\hat\gamma_0, \hat\gamma_1, \hat\gamma_2, \hat\gamma_3) = (0.45, 0.56, 1.07, 2.09)$.
All four estimates are positive, indicating that a larger proportion of irrigated area is associated with higher soybean yield in each of the four years after accounting for the temperature trajectory. Although the magnitude of this association varies across years, the direction is consistent, supporting the importance of water availability for soybean production in Kansas.
As with the functional coefficients, the magnitude of $\hat\gamma_k$ should not be interpreted as an ordinary linear-regression slope. If the irrigation predictor changes from $z$ to $z+\Delta z$, then, up to the corresponding normalizing constants, the conditional density is tilted by a factor $\exp\{\gamma_k\Delta z\,y\}$. Thus, a positive $\gamma_k$ indicates that greater irrigation shifts the conditional distribution toward larger yield values. The estimates $(0.45,0.56,1.07,2.09)$ therefore provide evidence of a consistently positive conditional association, while the differences in magnitude indicate that the strength of the distributional shift may vary across years.

The results should be interpreted as associations rather than causal effects. The analysis adjusts for the observed temperature trajectory and irrigation variable, but other annual environmental and management factors are not fully represented in the data. In particular, the humidity and sunshine information discussed in the Introduction is unavailable. The year-specific baseline distributions provide flexibility for unexplained distributional differences across years, but they do not remove confounding by unmeasured county-level factors. This limitation is important when interpreting the agricultural conclusions of the analysis.

\section{Conclusion and Discussion}
We proposed a semiparametric functional generalized linear model for data from multiple related populations. The proposed framework leaves the population-specific baseline response distributions unspecified and connects them through a density ratio model. This formulation reduces dependence on a fully specified parametric response distribution while allowing information to be shared across populations. We developed a maximum empirical likelihood estimation procedure, used penalized B-splines to estimate the functional coefficients, and introduced a cross-validation strategy for selecting the smoothing parameters.

The simulation studies illustrate two complementary advantages of the proposed DRM-EL approach. First, when multiple populations are related through the density ratio structure, jointly analyzing the samples can improve estimation efficiency compared with fitting the populations separately. Second, when a conventional parametric response model is misspecified, the semiparametric formulation can provide substantially more stable estimation because the baseline distributions are not forced into a prespecified family. The Kansas soybean application further demonstrates how the framework can be used to study population-specific relationships involving both scalar and functional predictors.

Conceptually, the proposed DRM formulation lies between two extreme analysis strategies. At one extreme, each population can be fitted separately, which requires few cross-population assumptions but may be inefficient when individual sample sizes are small. At the other extreme, all populations can be completely pooled under a common baseline distribution or common regression parameters, which can be efficient but may obscure meaningful heterogeneity. The proposed method shares information only through a structured relationship among the baseline distributions while allowing the regression coefficients to remain population-specific. This partial-pooling perspective is especially useful when samples arise from related years, sites, studies, or subpopulations for which complete homogeneity is implausible.

The efficiency gain from the DRM is not automatic and depends on whether the selected basis $\bm q(y)$ provides an adequate representation of the relationships among the baseline distributions. If the density-ratio specification is severely misspecified, borrowing information may introduce bias rather than improve estimation. For this reason, practical implementation should be accompanied by sensitivity analyses over reasonable choices of $\bm q(y)$ and, when possible, goodness-of-fit assessment for the density-ratio structure. The current simulation studies use correctly specified DRM bases and therefore primarily characterize performance when the population linkage is adequately modeled.

Several extensions warrant further investigation. The current development focuses on continuous responses; extending the framework to categorical, count, or other discrete outcomes would broaden its applicability. In addition, the soybean analysis incorporates the daily temperature trajectory through the linear functional index $\langle X,\beta_k\rangle$. Although $\beta_k(t)$ allows temperature to have positive effects at some times and negative effects at others, the model imposes a restriction: holding other predictors fixed, raising temperature by the same amount on every day can only increase, decrease, or leave unchanged the predicted mean yield. It cannot capture a pattern in which moderate warming improves yield but further warming reduces it, such as adverse effects of both excessively low and excessively high temperatures. Future work could accommodate such nonmonotone effects through functional additive models or functional single-index models with flexible, potentially nonmonotone link functions. Finally, because the density ratio framework borrows information across related populations through the basis $\bm q(y)$, data-adaptive selection of this basis is another direction for future research.

\section*{Disclosure statement}\label{disclosure-statement}
None declared.

\section*{SUPPLEMENTARY MATERIAL}

%\end{center}

Supplementary material is available online.
% \href{http://biostatistics.oxfordjournals.org}%
% {http://biostatistics.oxfordjournals.org}.

\bibliography{Bibliography-SFQR, References, bibfile}

\end{document}